\documentclass[twocolumn,pra]{revtex4}

\usepackage{graphicx}
\usepackage{dcolumn}
\usepackage{amsmath}

\makeatletter
\def\btt#1{\texttt{\@backslashchar#1}}%
\DeclareRobustCommand\bblash{\btt{\@backslashchar}}%
\makeatother

\begin{document}


\title[Short Title]{Characterization of high finesse mirrors: \\ loss, phase shifts and mode structure in an optical cavity}

\author{Christina J. Hood}
\author{H. J. Kimble}
 \affiliation{Norman Bridge Laboratory of Physics 12-33, California Institute of Technology, Pasadena, CA 91125}
\author{Jun Ye}
\affiliation{JILA, National Institute of Standards and Technology
and University of Colorado, Boulder, CO 80309}

\date{\today}

\begin{abstract}
An extensive characterization of high finesse optical cavities
used in cavity QED experiments is described. Different techniques
in the measurement of the loss and phase shifts associated with
the mirror coatings are discussed and their agreement shown.
Issues of cavity field mode structure supported by the dielectric
coatings are related to our effort to achieve the strongest
possible coupling between an atom and the cavity.
\end{abstract}

\pacs{Valid PACS appear here}%

\maketitle

\section{Introduction}

For many contemporary physics experiments, measurement enhancement
via an optical cavity is a useful tool. Indeed, an optical cavity
allows one to extend the interaction length between matter and
field, to build up the optical power, to maintain a well-defined
mode structure, and to study the extreme nonlinear optics and
quantum mechanics associated with the large field of a single
photon for small cavity volumes \cite{HJK1}.\ In most situations,
a better understanding of cavity and mirror properties is
important for achieving improved sensitivity and for elimination
of systematic errors. For example, in cavity QED, one needs to
know the mode structure of the intracavity field in order to
develop the optimum strategy of atom-cavity coupling; for
frequency metrology, accurate determination of phase shifts of
the resonant fields can provide precision frequency markers; and
in quantitative spectroscopy, knowledge of the mirror loss sets
the accuracy scale of absorption measurement. On the technology
development side, the knowledge gained from careful mirror
characterization could provide guidelines for the optic coating
community to develop in situ measurement and control capabilities
of the coating process.

The work presented in this paper is motivated by the
ever-increasing demand for a high coherent coupling rate between
an atom and the field, as well as of a decreasing cavity loss
rate. The aim is to have coherent (reversible) evolution
dominating over dissipative processes, and thereby to explore
manifestly quantum dynamics in real time, which in turn should
lead eventually to the investigation of the strong conditioning of
system evolution on measurement results and the realization of
quantum feedback control. An important feature associated with
strong coupling is that system dynamics are readily influenced by
single quanta. Thus single-atom and single-photon cavity QED
provides an ideal stage where the dynamical processes of
individual quantum systems can be isolated and manipulated. A
collection of such coherent systems could help to realize a
distributed quantum networks for computation and communication
\cite{ENK}. At each node the quantum information is stored by one
or a collection of entangled atoms. Photons serve as the
communication link which in turn entangle the whole network.
Within this context, technical advances in optical cavity quantum
electrodynamics have become increasingly important. Some
significant developments down this road have been achieved by the
group at Caltech \cite{HMabuchi96, Hood98, HMJY99, Ye99, Yeprl99,
Hood2000}, where in Ref. \cite{Hood2000} the one-photon Rabi
frequency is $\Omega _{1}/2\pi =220$MHz, in comparison with the
atomic decay rate $\gamma _{\bot }=2.6$MHz and the cavity decay
rate $\kappa /2\pi =14.2$MHz.

The strong coupling condition $\Omega _{1}\gg (\gamma _{\bot
},\kappa )$ is achieved by using a small cavity length, of the
order of 10$\mu $m. Precise measurement of the length of a short
optical cavity facilitates the determination of mirror coating
characteristics. A 10$\mu $m cavity length translates to a
free-spectral-range ($FSR$) of 15 THz, or a wavelength difference
of a few tens of nanometers (for example, it is 36 nm for a
center wavelength of 852 nm) for neighboring cavity modes.
Therefore a straightforward 6-digit measurement of the
wavelengths (Burleigh wavemeter) of the cavity modes acquires a
precision of the order of 5 x 10$^{-5}$ for accurate
determination of the equivalent optical length of the cavity, from
which details of the index of refraction and layer thickness of
materials in the mirror stack can be inferred.

The low loss rate of the cavity field is made possible by high
quality mirror coatings that lead to scatter and absorption
losses in the 10$^{-6}$ range \cite{Rempe92,REO}. The cavity
finesse and overall cavity transmission can be measured directly
to determine the mirror losses, $l$, and transmission, $T$. This
information can be combined with the $FSR$ measurement in two
useful ways: Firstly, the $FSR$ measurement is sensitive to the
difference in refractive index $n_{H}-n_{L}$ of the materials
making up the multilayer mirror stack, whereas the transmission
$T$ depends on the ratio $n_{H}/n_{L}$, as will be shown later.
As a result, a precise measurement of both the $FSR$ and $T$ can
be used to determine the values of $n_{H}$ and $n_{L}$
independently. Moreover, by mapping out the wavelength dependence
of the $FSR$, the thickness of layers in the mirror stack can be
determined. Secondly, if one of the refractive indices (here
$n_{L}$) is well known, then the $FSR$ measurement determines
$n_{H}$, and an independent value for the mirror transmission $T$
can then be calculated from $n_{H}$ and $n_{L}$, and compared to
the experimentally measured result. Indeed, the work presented in
this paper shows that we are able to make complementary and
mutually-confirming measurements of the cavity properties by the
two approaches, i.e., measurements of the direct cavity loss and
the dispersion of the cavity modes.

Coming back to the cavity QED experiments, we note that knowledge
of the cavity properties is of importance in two particular ways:
1. Mirror absorption/scatter losses are a critical limiting
factor in the loss rate from our cavity QED system - for our
current cavities the loss rate from photon scattering due to
mirror imperfections is similar in size to the atomic spontaneous
emission rate. To build robust quantum computing/communications
devices from cavity QED components, it is necessary to improve
the ratio of mirror transmission to mirror losses. 2. The
standing-wave light field inside the cavity penetrates into the
mirror coatings, giving a larger mode-volume $V_{\text{mode}}$
than would be expected naively from the physical distance between
the mirror surfaces. Since $\Omega _{1}\propto
1/\sqrt{V_{\text{mode}}}$, as our micro-cavities are pushed to
shorter lengths, this leakage field will have a non-negligible
effect on the achievable coupling strength $g_{0}=\Omega _{1}/2$.

\section{Direct Transmission and Loss Measurements}
\label{sec:seclossmeasurements}

All of the mirrors described in this paper were fabricated by
Research Electro-Optics in Boulder, Colorado\cite{REO}. More
specifically, the measurements were made for the particular
coating run REO \#T95 and involved mirrors with radius of
curvature R = 10 and 20 cm. The coating run had a design
transmission of $T^{th}$=7ppm at a center wavelength of 852nm,
from which a cavity finesse of $\mathcal{F}$=370,000 was
expected. It was somewhat surprising therefore to measure a
finesse of $\mathcal{F}$=480,000 at the targeted wavelength, and
this prompted us to make more detailed measurements of the mirror
properties, and design a model to match these measurements.

Firstly, losses were measured directly with a 40$\mu m$ length
cavity of 20cm radius of curvature mirrors in the usual way by
recording resonant cavity transmission, reflection and finesse.
If we denote the transmission of mirrors 1 and 2 by $T_{1}$ and
$T_{2}$, respectively, and the (absorption + scatter) loss per
mirror as $l_{i}=(A+S)_{i}$, then the total cavity losses
$\mathcal{L}=T_{1}+T_{2}+l_{1}+l_{2}$ can be determined from the
cavity finesse $\mathcal{F}=FSR/2\kappa $, with $FSR$ as the
cavity free spectral range and $\kappa $ as the HWHM for the
TEM$_{00}$ mode of the cavity; equivalently, $\mathcal{F}=2\pi
/\mathcal{L}$. The cavity linewidth $ \beta =2\kappa $ can be
determined from a ringdown measurement or using a modulation
sideband as a frequency marker with the cavity length scanned,
which is the technique employed here. The cavity transmission
$I_{trans}= \frac{4T_{1}T_{2}}{(T_{1}+T_{2}+l_{1}+l_{2})^{2}}$
can then be used to determine $l_{1}+l_{2}$, if $T_{1}$ and
$T_{2}$ are known independently. In practice this is a difficult
measurement to make, because the overall transmission $I_{trans}$
depends on the mode-matching into the cavity being perfect. A
variation of this protocol that does not require perfect
mode-matching can be derived, by comparing the cavity reflection
and transmission values with the cavity locked on resonance and
off resonance.

The rudiments of this protocol are as follows. First of all, the
total loss ($\mathcal{L}=T_{1}+T_{2}+l_{1}+l_{2}$) is always
measured first with the determination of the cavity $FSR$ and
linewidth. Now let us denote the input power as $P_{in}$, the
reflected power $P_{r}$ , and the transmitted power $P_{t}$.
There is also a mode matching factor $\epsilon $, meaning that of
the input power of $P_{in}$, only $\epsilon P_{in}$ is useful for
coupling to the cavity TEM$_{00}$ mode, $(1-\epsilon )P_{in}$ is
wasted. We have the following equations (the assumption of two
equal mirrors is reasonable since the two mirrors are produced in
the same coating run)

\begin{equation}
\mathcal{F}=\frac{2\pi }{T_{1}+T_{2}+l_{1}+l_{2}}=\frac{\pi
}{l+T} \label{coatingeq1}
\end{equation}

\begin{equation}
\frac{P_{t}}{\epsilon P_{in}}=4T_{1}T_{2}(\frac{\mathcal{F}}{2\pi
})^{2}=T^{2}(\frac{\mathcal{F}}{\pi })^{2} \label{coatingeq2}
\end{equation}

\begin{equation}
\frac{P_{r}-(1-\epsilon )P_{in}}{\epsilon
P_{in}}=(l_{1}+l_{2}+T_{1}-T_{2})^{2}(\frac{\mathcal{F}}{2\pi
})^{2}=l^{2}(\frac{\mathcal{F}}{\pi })^{2}
\label{coatingeq3}
\end{equation}

Remember that $(1-\epsilon )P_{in}$ is the ``useless'' power that
is reflected directly off of the input mirror, and must be
subtracted from $P_{r}$ to leave the reflected power we wish to
measure, that is, the sum of the field leaked from the cavity
storage and the field (mode-matched) directly reflected off the
input mirror. This cavity contrast is a direct result of the
mirror properties. Division of equation \ref{coatingeq2} by
\ref{coatingeq3} gives

\begin{equation}
\frac{P_{t}}{P_{r}-P_{in}}=\frac{T^{2}(\frac{\mathcal{F}}{\pi
})^{2}}{l^{2}(\frac{\mathcal{F}}{\pi })^{2}-1}\qquad
\label{coatingeq4}
\end{equation}
Equation \ref{coatingeq4}, combined with \ref{coatingeq1}, will
determine completely $T$ and $l$ .

In the actual experiment, this direct measurement approach found
that (from finesse we have $l+T=7.2ppm$) $P_{in}=54\mu W$,
$P_{r}=42.6\mu W$ and $P_{t}=4.82\mu W$ and therefore $l=2.9ppm$
and $T=4.3ppm$, with measurement uncertainties below $5\%$.

Another way to measure the ($T,l$) is by sweeping out all the
high order spatial modes and carefully noting the transmission
and reflection powers at each spatial mode. One measures the
total input power and also sums together the powers of every
matched mode for transmission and reflection. These three powers
can be used in Eq. \ref{coatingeq2} and \ref{coatingeq3} to
calculate the partition between $T$ and $l$. That measurement produced $%
l=3ppm$, and $T=4.2ppm$. The value of $T$ should be a bit lower
in this case because it is not possible to include all higher
order modes in the measurement, some of them are simply
impossible to resolve due to their weakness.

Other cavities measured with mirrors from the same coating run
had higher finesse, very likely due to a lower density of surface
defects. To construct a cavity of minimal mode volume for the
intended maximal coherent coupling rate, we need to have the
distance between two mirrors (radius of curvature $R = 10 cm$) on
the order of $10 \mu m$ or below. To avoid contact between the
outer edges of the two mirrors, the mirrors were fabricated with
cone-shaped fronts, reducing the substrate radius from 3 mm to 1
mm. We notice this extra machine process might have introduced
some additional surface defects on some mirrors. However, the
highest finesse achieved with cone-shaped mirrors was comparable
to unmodified pieces, at $\mathcal{F}$=480,000$\pm $10,000,
corresponding to losses $l=2.2ppm$ if mirror transmission
$T=4.3ppm$ as determined from the above measurements.

\section{Technical details of the model}

In this section we derive a model for the coating properties. A
transfer-matrix formalism was used to calculate the input-output
propagation of a plane-wave field through the 37 layer stack of
alternating high index (Ta$_{2}$0$_{5},$ $n_{H}$=2.0411) and low
index (SiO$_{2},$ $n_{L}$=1.455) dielectric layers (these
dielectric constants are assumed to be constant with wavelength).
The substrate refractive index (supplied by REO)\ used was
$n_{sub}=$1.5098. That is, the transfer of the field through each
$\lambda /4 $ layer is represented by a matrix, and the response
of the entire mirror (or cavity)\ is determined by the product of
these individual matrices.

Following the treatment of Hecht \cite{hecht} for normal
incidence, we take the matrix representing layer $j$ to be given
by M$_{j}=\left[
\begin{array}{cc}
\cos (kh_{j}) & (i\sin (kh_{j}))/Y_{j} \\
iY_{j}\sin (kh_{j}) & \cos (kh_{j})
\end{array}
\right] .$ Here M$_{j}$ relates the electric and magnetic fields
$(E,H)$ of the input and output via

\begin{equation}
\left[
\begin{array}{c}
E_{out} \\
H_{out}
\end{array}
\right] =\left[ M\right] \left[
\begin{array}{c}
E_{in} \\
H_{in}
\end{array}
\right] \text{.}
\end{equation}
$k=2\pi /\lambda $ is the free-space wavevector of the incident
light, $h_{j}=n_{j}$ x (layer thickness) with $n_{j}$ the
refractive index, and $Y_{j}=\sqrt{\frac{\epsilon _{0}}{\mu
_{0}}}n_{j}$ with ($\epsilon _{0},\mu _{0}$) the electric and
magnetic constants in SI units. For an exact $\lambda /4$ layer
(and for light at the design wavelength of the coating), this
simplifies to M$_{j}=\left[
\begin{array}{cc}
0 & i/Y_{j} \\
iY_{j} & 0
\end{array}
\right].$ A multilayer stack is represented by multiplying the
matrices of the individual layers:\ For light incident on layer
1, the matrix for the entire structure of q layers is defined as
the product $M=M_{1}M_{2}...M_{q}. $ For our mirror stack, this
gives $M=(M_{Ta_{2}O_{5}}M_{SiO_{2}})^{18}M_{Ta_{2}O_{5}}.$ Note
that at the coating center (where there is an exact $\lambda /4$
layer), $M_{Ta_{2}O_{5}}M_{SiO_{2}}=\left[
\begin{array}{cc}
-\frac{n_{L}}{n_{H}} & 0 \\
0 & -\frac{n_{H}}{n_{L}}
\end{array}
\right] $, so the system matrix has the simple form $M=\left[
\begin{array}{cc}
0 & \frac{i}{Y_{H}}(\frac{n_{L}}{n_{H}})^{18} \\
iY_{H}(\frac{n_{H}}{n_{L}})^{18} & 0
\end{array}
\right] .$

For a field incident from material with index $n_{0}$ and exiting
into material with index $n_{s},$ the resulting transmission
coefficient is given by

\begin{equation}
\ t=2Y_{0}/(Y_{0}M_{11}+Y_{0}Y_{S}M_{12}+M_{21}+Y_{S}M_{22}),
\end{equation}
with transmission $T=\frac{n_{s}}{n_{0}}|t|^{2}$ (the factor
$\frac{n_{s}}{n_{0}}$ accounts for the change in the amplitude of
the electric field in the dielectric, thereby conserving the net
energy flux). At the center wavelength of the coating then,

\begin{equation}
\ T=\frac{n_{s}}{n_{0}}%
|-2i/[(n_{S}/n_{H})(n_{L}/n_{H})^{18}+(n_{H}/n_{0})(n_{H}/n_{L})^{18}]|^{2}.
\end{equation}
We can make a further simplification: as
$(n_{L}/n_{H})^{18}=0.0018$ and $(n_{H}/n_{L})^{18}$ =557, the
first term in the denominator of the above equation is only a
10$^{-6}$ correction, so the final result for $T$\ at the coating
center becomes

\begin{equation}
T=4n_{S}n_{0}(n_{L})^{36}/(n_{H})^{38},
\end{equation}
and the transmission is determined by the $ratio$ of the
refractive indices.

\begin{figure}[bp]
\vspace{0.5in}
\resizebox{7.5 cm}{!}{
\includegraphics{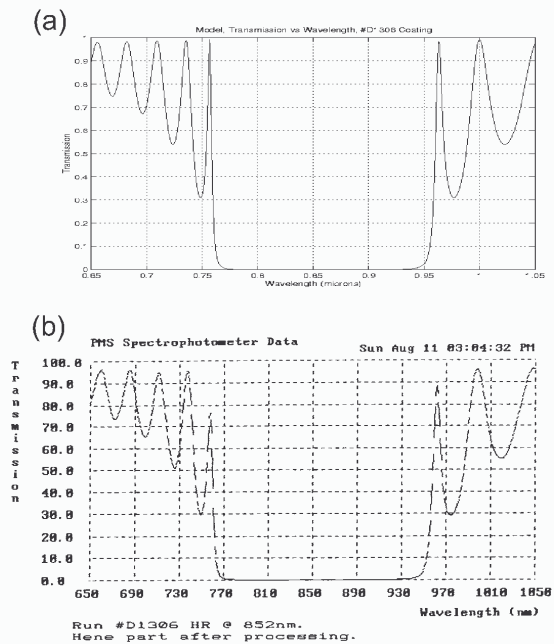}}
\caption{(a) Calculated and (b) Measured transmission of coating
as a function of wavelength, for a 35 layer $\protect\lambda /4$
stack with $n_{H} $=2.0411, $n_{L}$=1.455, and center wavelength
850nm.} \label{Fig1}
\end{figure}

This calculation reproduced the target reflectivity of
$T^{th}$=7.3ppm for the coating run \#T95, and $T^{th}$=14.6ppm
for another REO coating run \#D1306 where the number of layers
was reduced to 35. The model and measured (REO spectrophotometer
data) ``coating curves'' are shown in Fig. \ref{Fig1} for the
\#D1306 coating run.

For a fixed cavity length the resonance wavelengths of the cavity
can be calculated simply with the same transfer-matrix formalism,
using a matrix for the entire system, $M_{total}=MM_{gap}M$, (a
product of two mirrors plus a fixed-length vacuum gap in
between). The calculation steps through a series of wavelengths
calculating the cavity transmission $T$ at each, and by finding
places of maximum transmission finds the vacuum wavelengths of
the cavity resonances.

Conversely, for a given set of measured cavity-resonance
wavelengths, it is possible to determine the effective cavity
length precisely. With a commercial wavelength-meter that gives
6-digit wavelength measurement, we typically measure the cavity
resonance within an uncertainty of 0.01 nm. Error propagation
analysis gives an uncertainty for the determination of the
effective cavity length (tens of microns) on the order of $0.05 -
0.1$ nm. The parameters of the model (index contrast, layer
thickness)\ are set by comparison to such measurements. Hence,
armed with the detailed knowledge of the mirrors provided from
the model, the physical cavity length can be determined precisely
from a single measurement of resonance, for example, when the
cavity is locked to a laser of known frequency (in our case a
cesium transition at 852.359nm). Close to the center of the
design wavelength of the coating, the effective cavity length (on
resonance) is roughly $L_{eff}=L\ +1.633\lambda /2$ with $L$ (the
physical distance between the mirror surfaces)\ an integer number
of $\lambda /2$. The physical cavity length can therefore be
determined, with an uncertainty of $\sim$ 0.5 nm, limited by the
overall parameter-fitting in the model. Further details of the
wavelength-dependence are provided by reference to the model.

\section{Free-Spectral Range (FSR) Measurements}

To determine the parameters of the model (index contrast, layer
thickness), a series of precise measurements of the cavity FSR
(frequency between successive cavity resonances)\ was
made\cite{devoe,lichten,layer}. At fixed cavity length a
Ti-Sapphire laser was tuned to find successive resonant
wavelengths ($\lambda _{1},\lambda _{2}$) of the cavity, and an
experimentally determined length was then defined by
$L_{expt}=\lambda _{1}\lambda _{2}/2(\lambda _{1}-\lambda _{2}).$

This length comprises the actual physical length between the two
mirror surfaces, $L$, plus a contribution from leakage of the
mode into the mirror stack, which gives rise to an additional
phase shift at the coatings, to give a length $L_{eff}>L.$ In
addition, the leakage into the coatings increases with wavelength
as ($\lambda _{1},\lambda _{2}$) move away from the coating
design wavelength, so this gives another additional contribution
to the round-trip phase and hence to the measured length
$L_{expt}.$

As discussed in Ref. \cite{devoe}, if $\lambda _{1}$ and $\lambda
_{2}$ were closely spaced compared to the scale on which the
coating properties vary (so that coating dispersion could be
neglected) then near the design wavelength of the coating, we
would have $L_{expt}=L_{eff}=L\ +\ (\frac{1}{n_{H}-n_{L}})\times
\lambda _{c}/2$ where $n_{H}$ and $n_{L}$\ are the high and low
index materials of the stack, and $\lambda _{c}=2\lambda
_{1}\lambda _{2}/(\lambda _{1}+\lambda _{2})$ is the average (in
frequency)\ of wavelengths $\lambda _{1}$ and $\lambda _{2}$. We
thereby have a dependence of the free spectral range on
$(\frac{1}{n_{H}-n_{L}}),$ which combined with the transmission
(which depends on $n_{L}/n_{H}$) can fix $ n_{H}$ and $n_{L}.$
For these materials, this gives $L_{eff}=L\ +1.633\lambda
_{c}/2.$ However, for our measurements with short cavities,
$\lambda _{1} $ and $\lambda _{2}$ are separated by $\simeq
$30nm, so $L_{expt}>L_{eff}$. But we can still use the complete
model to fit to the measured values ($\lambda _{1},\lambda _{2}$)
and determine parameters of the coating. Finally, by mapping out
this wavelength dependence of the free-spectral range to find
$min(L_{expt})$, we find the center wavelength of the coating.

In the model, the refractive indices used are adjusted to obtain
the same pairs ($\lambda _{1},\lambda _{2}$) as measured. Then,
the layer thickness in the model is adjusted to agree with the
measured coating center wavelength. By using the additional
information of the measured mirror transmission $T$ from Section
\ref{sec:seclossmeasurements}, we can now either:

1. Derive independent values for the refractive indices and layer
thickness, or

2. Assuming one index is known, use the refractive indices and
layer thickness information to give an independent value for the
mirror transmission, which can be compared to the measurement of
Section \ref{sec:seclossmeasurements}.

That the dispersion (FSR) measurement alone is sufficient to
determine the loss-less part of the mirror properties represents
some useful information for the mirror coating technician: the
index difference $n_{H}-n_{L}$ \ and the optical thickness of the
coating layers can be simply measured in this way without
interference from absorption/scatter losses. And, if $n_{L}$ is
known, this also gives a simple way of finding the mirror
transmission. Adding in a direct measurement of mirror
transmission yields values for $n_{H}$ and $n_{L}$ separately.

\begin{figure}[bp]
\resizebox{7.5cm}{!}{
\includegraphics{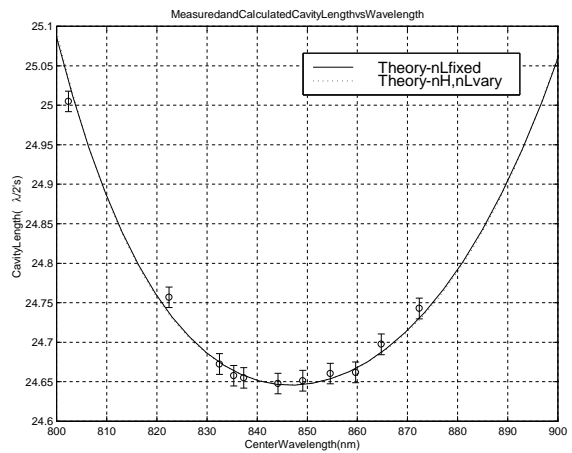}}
\vspace{0.25cm} \caption{The cavity length $L_{expt}$ measured
from the Free-Spectral Range (FSR)\ varies about the design
wavelength of the coating. Fitting a model to these data points
gives a measure of mirror transmission (from fitting of the
difference $n_{H}-n_{L}$) and center wavelength (from fitting
layer thickness).} \label{Fig2}
\end{figure}

Data obtained from these measurements are shown in Figure
\ref{Fig2}, where $L_{expt}$ is plotted as a function of
wavelength, for a 10 $\mu $m cavity with 10 cm radius of
curvature mirrors. The circles are measured data, and the curves
the calculation from the model, with parameters chosen to best
fit the data. This data was taken by setting the cavity to a
series of different lengths, and recording a pair of resonant
wavelengths $(\lambda _{1},\lambda _{2})$ at each length. The x
axis is center wavelength $\lambda _{c}=2\lambda _{1}\lambda
_{2}/(\lambda _{1}+\lambda _{2})$, the y axis the measured cavity
length $L_{expt}=\lambda _{1}\lambda _{2}/2(\lambda _{1}-\lambda
_{2})$ shown in units of $\lambda _{1}/2$ :\ for each pair $
(\lambda _{1},\lambda _{2})$, the length is such that
$L_{expt}/(\lambda _{1}/2)=24.$xx . Dividing by $\lambda _{2}$
instead would exactly give 23.xx, since by rearranging the
formula for $L_{expt}$ we see that $ L_{expt}/(\lambda
_{1}/2)\equiv L_{expt}/(\lambda _{2}/2)+1$. Due to a finite drift
in the cavity length, each measurement of $\lambda $ was made to
only 5 digits resolution (e.g. 852.59$\pm $0.01nm), leading to the
uncertainty in $L_{expt}$ shown. Uncertainty in $\lambda _{c}$ is
$\pm $ 0.03nm and cannot be seen on this scale.

Two theory curves are shown. The solid curve shows a model with
$n_{L}$ assumed to be fixed at its nominal value of
$n_{L}$=1.455. To best fit the data, $n_{H}$ was $increased$ to
$n_{H}$=2.0676 (a factor of 1.3\%). In addition, the center
wavelength was shifted to 847nm (by reducing the thickness of
each $\lambda /4$ layer by 0.6\%). Discussions with REO confirmed
that 1.3\%\ is a known offset in  $n_{H}$ for the particular
coating machine which produced this run, and also that a few nm
uncertainty in the center wavelength is typical. With these
parameters, the inferred mirror transmission is $T_{\inf
}=$4.6$\pm $0.2 ppm, agreeing well with the measured value\
$T_{\exp }=$4.3 ppm\ from Section \ref{sec:seclossmeasurements}.
The dotted curve (which overlaps the solid curve) shows the model
when both $n_{L}$ and $n_{H}$ are allowed to vary. Their values
are chosen to match both the FSR\ measurement shown, and to give a
mirror transmission to match exactly the experimentally
determined value $T_{\exp }=4.3$ ppm. Parameters which satisfy
these criteria are $n_{H}$ =2.0564 (0.75\% increase)\ and
$n_{L}$=1.4440 (0.76\%\ decrease). Our direct measurement of $T$\
in Section \ref{sec:seclossmeasurements} had a large uncertainty,
which limits the absolute determination of $n_{H}$ and $n_{L}$ to
about this 1\%\ level. However, a more precise measurement could
in principle determine the indices at the 0.1\%\ level. One
application might be to measure $T$ and the FSR\ as a function of
position across a mirror substrate, thereby mapping out stress
induced variations in the refractive indices at the 0.1\%\ level
with a spatial resolution of $\sim $ 10 $\mu $m.

In this data set the correction for the Gaussian phase difference
between the actual resonator mode and the plane-wave of the model
has been neglected. After the propagation distance from the mode
waist to the mirror surfaces, a Gaussian beam will have acquired
less phase than a plane wave traveling the same distance. For a
10 $\mu $m cavity with 10 cm radius of curvature mirrors, this
gives a 2\%\ correction, corresponding to a shift in $L_{expt}$
by $\simeq 0.0045$ cavity orders (that is, $\Delta L\simeq +\frac{
\lambda }{2}\times 0.0045$). Lowering the refractive index
contrast of the model to shift the calculated curve by this
amount would increase the inferred mirror transmission by $
\lesssim $ 0.1 ppm. For our second cavity (44 $\mu $m, 20 cm
radius of curvature mirrors), the correction is 0.0066 cavity
orders.

\vspace{0.25cm}

The mirror phase shift (FSR\ measurement)\ is only sensitive to
the transmission (index contrast)\ and center wavelength (layer
thickness). Therefore, if absorption/scatter losses are added to
the model (by introducing an imaginary component to the
refractive index)\ the cavity resonance wavelengths do not
change. More precisely, adding a scattering loss at the mirror
surfaces has exactly zero effect on the FSR\ and mirror
transmission. Adding losses\textit{\ within }the coatings has a
small effect:\ increasing the mirror absorption from 0.5ppm to
2ppm (an experimentally reasonable range)\ changes the mirror
transmission by a factor of $\simeq 10^{-5}T$, clearly
negligible, and again there is no effect on the FSR\ measurement.
As a result, this measurement (with $n_{L}$\ assumed fixed)
provided a very simple and sensitive inference of the mirror
transmission of $T_{\inf }=$4.6$\pm $0.2 ppm, which is
\textit{unaffected} by absorption/scatter losses.

The same measurement and fitting procedure was used on another
cavity with mirrors from the same coating run. This 44 $\mu $m
cavity made from 20 cm radius of curvature mirrors gave a
transmission of $T_{\inf }=$4.5$\pm $0.2 ppm, with a center
wavelength of 848 nm.(This was the cavity used for the direct
measurements of Section \ref{sec:seclossmeasurements} which gave
$T$=4.3 ppm).

One other factor which has been ignored so far is the effect of
fluctuations in the $\lambda /4$ layer thickness. Discussions
with REO\ suggested that a 1\%\ variation in thickness was
reasonable, so a Gaussian-distributed variation (of standard
deviation 1\%)\ was added to the layer thicknesses of the model.
For cavity calculations, identical mirrors were used for both
sides of the cavity. The principal effect of this variation is to
shift the center wavelength of the coating - over several
realizations of random coatings, this resulted in an rms shift of
the center wavelength by $\pm 1.2$ nm. So, the measured shift of
center wavelength in the coating (from 852 nm to 847 nm)\ is
probably due partly to a systematic offset, and partly to
fluctuations. The mirror transmission is also affected: the value
of the transmission is on average\textit{\ increased
}slightly\textit{,} by 0.6\%\ in the case studied, from 4.55 ppm
to 4.58 ppm at the center of the coating. At the level of our
current measurements this is another negligible effect, but with
a more precise measurement aimed at determining $n_{H}$ and
$n_{L},$ the possibility of a systematic offset from this
mechanism should be considered. Lastly, the FSR\ measurement is
mostly effected via the change in center wavelength of the
coating - the value of $min(L_{expt(simulated)})$ has a mean the
same as without the added fluctuations, and varies by only 0.0014
mode orders rms, again negligible for our purposes.

Another useful result of these calculations is that the free
spectral range of the cavity is well known, so that resonant
wavelengths of the cavity can be accurately predicted. This is
important for choosing a diode laser of correct center wavelength
to match the mode, for applications such as cavity locking or
dipole-force traps. With the idea of using a laser of $>$920 nm
wavelength to form an intracavity dipole-force
trap,\cite{kimbleICOLS} this knowledge was particularly
important: our Ti:Sapphire laser tuned only as high as 890nm so
cavity resonances in this wavelength range could not be measured,
only predicted. With the parameters chosen above for the model,
the following theoretical and experimental resonance wavelengths
resulted:

\begin{tabular}{llllll}
theory & 787.208nm & 818.659 & 853.255 & 890.798 & 930.683 \\
experiment & 787.170nm & 818.651 & 853.255 & 890.800 & \ \ \ \ N/A
\end{tabular}

The experimental value for the cavity resonance can then
confidently be predicted to be 930.7$\pm $0.05nm, and a diode
laser chosen accordingly.

\section{Limitations to Mode Volume}

\begin{figure}[bp]
\resizebox{7.5cm}{!}{
\includegraphics{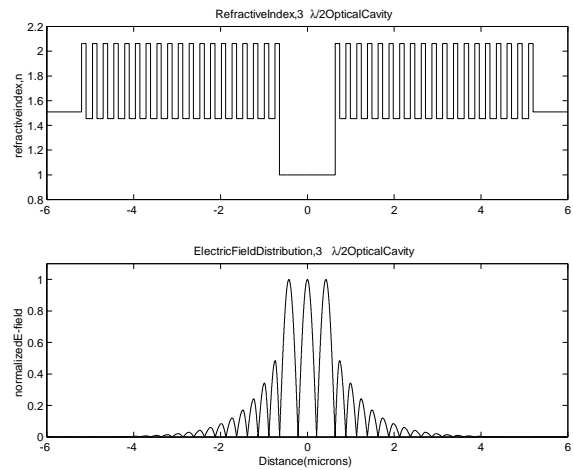}}
\caption{Mirror refractive index stack design, and resulting
electric field distribution for a resonant 3$\protect\lambda $/2
cavity made from dielectric mirrors.} \label{Fig3}
\end{figure}

In a similar calculation to the one described above, it is
possible to calculate the field distribution of light inside the
resonant cavity, by describing each layer separately with a left
and right travelling plane wave, then matching electromagnetic
boundary conditions between layers. An example of this kind of
calculation is shown in Figure \ref{Fig3},\ where refractive
index and field distribution (modulus of the electric-field) are
plotted as a function of distance for a cavity with length
$L_{eff}=3\lambda /2$. The coupling strength $g_{0}$ of an atom
placed in the center of the cavity mode is proportional to
$\frac{1}{\sqrt{V_{m}}}$, where $V_{m}$ the cavity mode volume is
found by integrating the field ($D\cdot E$)\ over the standing
wave and Gaussian transverse mode profile. Large coupling is
achieved by making a short cavity with a small mode waist (short
radius of curvature mirrors).

For a cavity of physical length $L$, the ``leakage'' of the mode
into the $\lambda /4$ mirror stack (look at the tails of the mode
in Fig. \ref{Fig3}) that increases $L$ to $L_{eff}$ also
increases the cavity mode volume. For our materials at 852nm,
$L_{eff}=L\ +1.633\lambda /2$ , so for a cavity with physical
distance between mirror surfaces $L=\lambda /2$, the cavity mode
volume ends up being 2.63 times larger than might otherwise have
been expected, and hence the atom-cavity coupling $g_{0}$ is 0.6
times smaller than the naive estimate based on the physical
separation of the mirror surfaces.

\begin{figure}[bp]
\resizebox{7.5cm}{!}{
\includegraphics{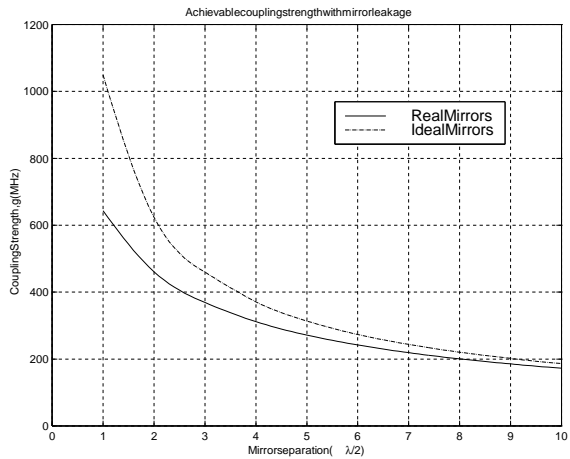}}
\caption{Coupling coefficient $g_{0}$ (expressed in cycles per
second) versus the physical separation $L$ of the surfaces of two
mirrors forming a Fabry-Perot resonator. Due to penetration of
the standing-wave mode into the mirror coatings, the cavity mode
volume achieved with real mirrors is larger (and hence the
coupling strength smaller)\ than for an ideal mirror with the
same spacing between mirror surfaces but no penetration.}
\label{Fig4}
\end{figure}

This effect is proportionately larger as the cavity length gets
shorter. In Figure \ref{Fig4}, the expected $g_{0}$ is plotted
for a cavity formed with two 20cm radius of curvature mirrors, as
a function of the physical distance $L$ between the mirrors. The
two curves show a real mirror (with $g_{0}$ reduced by leakage
into the coatings)\ and an idealized mirror with no leakage
(perfect reflectors at $\pm L/2$). The transverse (Gaussian waist)
dimension is calculated by simple Gaussian beam propagation,
which is not strictly accurate for length scales less than a few
microns; however any error in this should be roughly the same for
both the ideal and actual mirror cases, so the ratio of these
should remain sensibly correct. The cavity is assumed resonant at
an integer number of half-wavelengths of light at the 852nm Cs D2
transition; that is, each $\lambda /2$ is a distance of 0.426
microns.

The discrepancy between the expected and achieved coupling $g$ is
large even for our longer cavities - 5\%\ for a 10$\mu $m cavity.
However, in the lab this is largely compensated by the fact that
we never measure the actual physical distance $L$ between mirror
surfaces, but instead $L_{expt}=\lambda _{1}\lambda
_{2}/2(\lambda _{1}-\lambda _{2})$, which is close to $L_{eff},$
and so incorporates the same offset of mirror penetration that
determines $g_{0}$. This method of length measurement breaks down
eventually due to the dispersion of the mirror coatings:\
Eventually if $\lambda _{1}$ is at the center of the coating,
$\lambda _{2}$ will be so far separated in wavelength that it
reaches the edges of the mirror coating stopband, and the observed
round-trip phase has then more to do with the structure of the
dielectric coatings than it does with the vacuum gap between the
surfaces of the cavity mirrors. That is to say, our measured
$L_{expt}$ becomes increasingly different from $L_{eff}$ and
introduces an offset in estimating the mode volume as the cavity
length approaches the scale of the wavelength.

At $L=20\lambda /2$ physical length (the regime of our present
cavities) the difference between the coupling coefficient $g_{0}$
inferred from $L_{expt}$ and that found by integrating $D\cdot E$
over the mode volume is $<$0.1\%. At $L=10\lambda /2$ (4.26$\mu
$m)\ it would be a 1\% error, at $5\lambda /2$ an 8\% error. Note
however that knowledge of these offsets means that when
calculating $g_{0}$ from $L_{expt}$\ we can compensate for this
effect. Measurements of $L_{expt}$\ for cavities any shorter than
$5\lambda /2$ would be impossible since $\lambda _{2}$ has
reached the edge of the mirror stopband. To align shorter
cavities a new method for length measurement will need to be
developed, such as measuring the frequency spacing of transverse
modes.

We are now in a position to estimate parameters for the best
Fabry-Perot cavity that will be experimentally feasible in the
near future using this type of mirrors. First consider a\
$L=\lambda /2$ cavity with 20cm radius of curvature mirrors. If
the mirror transmission and losses were each reduced to
$T=l=$0.5ppm to yield a cavity finesse of \
$\mathcal{F}=3.14\times 10^{6} $ , then this cavity has
parameters $(g_{0},\kappa ,\gamma _{\bot })/2\pi
=(647,56,2.6)$MHz, which gives critical photon number
$n_{0}=\gamma _{\bot }^{2}/2g_{0}^{2}=8.1\times 10^{-6}$ and
critical atom number $N_{0}=2\kappa \gamma _{\bot
}/g_{0}^{2}=7.0\times 10^{-4}.$ To make a cavity of this length
the 20cm mirrors would have to be reduced to a diameter of 0.5mm
rather than 1mm. At this size there would still be a 0.11$\mu $m
gap between the mirror edges for the\ $L=\lambda /2$ (0.426
micron)\ cavity length, which should make it possible to still
get atoms into and out of the cavity (as in Ref.
\cite{HMabuchi96, Hood98, HMJY99, Ye99, Yeprl99, Hood2000}), and
to align the mirrors.

If the mirror diameter could be reduced to 350$\mu $m (without
adversely affecting the cavity losses), then 10cm radius of
curvature mirrors could be used, with a 0.12$\mu $m gap at the
edges. Due to the tighter radius of curvature, $g_{0}/2\pi $
would be increased to 770MHz in this case. Now speculating that
``dream''\ mirrors of $T=$0.2ppm transmission, $l=$0.2ppm loss
might be possible $(\mathcal{F}=7.85\times 10^{6})$, we could aim
for the ultimate goal of $(g_{0},\kappa ,\gamma _{\bot })/2\pi
=(770,22,2.6)$ MHz, in which case $n_{0}=5.7\times 10^{-6}$
photons, and $N_{0}=1.9\times 10^{-4}$ atoms.

In conclusion, we have presented two measurement approaches, one
based upon direct loss and the other on cavity dispersion, that
produce the same quantitative determination of the mirror coating
properties. The dispersion measurement is more informative, as it
has the potential to determine the complete characteristics of a
mirror. A model has been derived to link the mirror properties to
the physical parameters of coating layers. Issues relevant to
optical cavity QED, such as the cavity field mode structure, have
been discussed.

First and foremost, we thank Research Electro-Optics, Inc. for
providing the best quality mirrors and coatings that have made
our work possible. In particular, the critical and ongoing
contributions of R. Lalezari and J. Sandburg to our reserach
program in cavity QED are gratefully acknowledged. Our Caltech
colleagues David Vernooy and Theresa Lynn made important
contributions to the work presented here. Funding has been
provided in part by DARPA through the QUIC (Quantum Information
and Computing) program administered by the US Army Research
Office, the Office of Naval Research, the National Science
Foundation, and the National Institute of Standards and
Technology.

\end{document}